\documentclass[prb,superscriptaddress,showpacs]{revtex4}
\usepackage{graphicx}
\newcommand{\be}{\begin{equation}}
\newcommand{\ee}{\end{equation}}
\renewcommand\vec[1]{\mbox{\boldmath $#1$}}
\def\p{^{\,\prime}}

\newcounter{appendixc}
\renewcommand{\appendix}[1]{\vspace*{0.6cm}
\refstepcounter{appendixc} \setcounter{table}{0}
\setcounter{equation}{0}
\renewcommand{\thetable}{\Alph{appendixc}\arabic{table}}
\renewcommand{\theappendixc}{\Alph{appendixc}}
\renewcommand{\theequation}{\Alph{appendixc}\arabic{equation}}
\noindent{\bf Appendix \theappendixc. #1}\par\vspace*{0.4cm}}

\begin{document}

\title{Dispersion of confined optical phonons in semiconductor nanowires in the framework of a
continuum approach}
\author{F. Comas}
\thanks{On leave from: Departamento de F\'{\i}sica Te\'{o}rica, Universidad
de la Habana, Vedado 10400, Havana, Cuba}
\affiliation{Departamento de F\'{\i}sica, Universidade Federal de
S\~ao Carlos, 13565-905 S\~ao Carlos SP, Brazil}
\author{I. Camps}
\affiliation{Departamento de F\'{\i}sica, Universidade Federal de
S\~ao Carlos, 13565-905 S\~ao Carlos SP, Brazil}
\author{G. E. Marques}
\affiliation{Departamento de F\'{\i}sica, Universidade Federal de
S\~ao Carlos, 13565-905 S\~ao Carlos SP, Brazil}
\author{N. Studart}
\affiliation{Departamento de F\'{\i}sica, Universidade Federal de
S\~ao Carlos, 13565-905 S\~ao Carlos SP, Brazil}
\date{\today}

\begin{abstract}
Confined optical phonons are discussed for a semiconductor nanowire of the Ge (Si)
prototype on the basis of a theory developed some years ago. In the present work this
theory is adapted to a non polar material and generalized to the case when the phonon
dispersion law involves both linear and quadratic terms in the wave vector. The treatment
is considered along the lines of a continuous medium model and leads to a system of coupled
differential equations describing oscillations of mixed nature. The nanowire is modelled
in the form of an infinite circular cylinder and the solutions of the fundamental
equations are found. We are thus led to a description of long wavelength optical phonons,
which should show a closer agreement with experimental data and with calculations along
atomistic models. The presented theory is applied to the calculation of optical phonons
in a Ge nanowire. We have found the dispersion curves for various optical
phonon modes. We also normalize the modes and discuss the electron-phonon interaction within
the deformation potential approximation.
\end{abstract}

\pacs{63.22.+m, 63.20.Dj,63.20.Kr}
\maketitle

\section{Introduction}

The study of semiconductor nanowires is of current importance both
for device applications and fundamental physics, a point that has
been recognized since a certain time ago.\cite{r1} More recently,
the fabrication of Si and Ge nanowires by means of different
techniques has been stimulated in view of their relevance in device
nanotechnology~.\cite{r2,r3,r4,r5,r6} The physical properties of
these nanostructures are being studied both theoretically and
experimentally, while the fabrication of field-effect
transistors on the basis of this type of systems has been also reported.
\cite{r7,r8,r9,r10,r11,r12,r13} In the present work we
are interested in the phonon properties of non polar nanowires of
the Ge and Si prototypes, and, particularly, in the optical
phonons. It is well-known that phonon properties play an important
role in the considered systems because of their significance for the
analysis of various physical processes, such as, charge and thermal
transport, optical transitions, etc. Moreover, phononic engineering
has been invoked for device application purposes,\cite{r14} making
it relevant to have a better description of phonon modes.

The analysis of phonon properties in quantum-wires already has a
long history. By applying the long-wavelength approximation, based on
continuum approaches, the polar optical phonons were discussed by
different authors more than a decade ago\cite{r15,r16,r17,r18}
(see also Ref.~\onlinecite{r19} for a more recent paper on the
subject). Considering the nanowires made of non polar materials,
we must remark the existence of a close related subject: phonons
in carbon nanotubes,\cite{r20,r21} an issue revealing outstanding
possibilities for present day nanotechnology.

In the case of nanowires of the Ge or Si prototypes the
calculation of optical phonons by means of long-wavelength continuum
approaches involves interest by itself, and also for applications.
In the framework of the present paper we shall support
calculations of the optical phonon modes on a phenomenological
theory, which was proposed more than a decade ago for the polar
optical phonons in the same kind of
nanostructures.\cite{r22,r23,r24} Using this approach polar
optical phonons in quantum-wires were studied in Refs. \onlinecite{r17,r18}.
In the Ref.~\onlinecite{r21} the same theory, but adapted to non
polar optical phonons, was applied to the case of carbon
nanotubes. In all the above mentioned works it is explicitly
assumed that the optical phonons display a dispersion law
quadratic with the wave vector, which, in the bulk semiconductor
case is of the form $\omega^2=\omega_0^2-\beta^2k^2$ and,
according to this long-wavelength approximation, should be considered
near the $\Gamma$-point of the Brillouin Zone (BZ).\cite{rr1} In
the present paper the theory is generalized in order to take
account of linear terms in the phonon dispersion law, i.e., we
assume a more general dispersion law having the form
$\omega^2=\omega_0^2-\beta\p k-\beta^2k^2$ (in the bulk case).
Hence, we propose more general equations that in
Ref.~\onlinecite{r21}, now involving linear terms, and also we
show how they can be solved. It is easy to realize that the
additional linear term allows a better description of the optical
phonons near the BZ center, in closer correspondence with both
experimental evidence and atomistic calculations involving
discretized microscopic models. In some of the studied cases this
linear term becomes essential, and this situation is actually realized
in nanostructures based on Si and Ge.

Then, the paper is organized as follows. In Section II we present the fundamental equations, discuss their physical meaning and other details of the approach. In Section III we show how the equations can be solved in the case of a wire of circular cross-section and infinite length. Section IV is devoted to a physical discussion of the results obtained, while the case of a Ge nanowire is addressed.

\section{Fundamental equations}

We assume that the relative displacement vector $\vec{u}$ satisfies the following equation:

\be \label{1}
\rho (\omega^2-\omega_0^2)\vec{u}=\rho\beta_L^2\nabla\nabla\cdot\vec{u}-\rho\beta_T^2\nabla\times\nabla\times\vec{u}+i\rho(\vec{a}\cdot\nabla)\vec{u}+i\rho\vec{b}(\nabla\cdot\vec{u})\,\,.
\ee

Equation~(\ref{1}) is analogous to that introduced in
Refs.~\onlinecite{r22,r23,r24,r17,r18}, just considering that the
electric potential is zero (see also Ref.~\onlinecite{r21}).
However, new terms were added (two last terms at the r.h.s. of
(\ref{1})), which modify the nature of the differential equation
and lead to the linear terms in the phonon dispersion law. The
parameter $\rho$ is the reduced mass density of the two atoms (Ge)
in the unit cell,  $\omega_0$  is the optical phonon frequency at
the $\Gamma$ point of the bulk semiconductor, and
$\vec{u}(\vec{r},\,t)$ is considered to depend harmonically on the
time ($\sim \exp (-i\omega t)$). Parameters $\beta_L$ and
$\beta_T$ (with dimensions of velocity) are introduced to describe
the quadratic (parabolic) phonon dispersion and used to fit the
bulk semiconductor dispersion curve. They were already considered
in previous works. Now we are introducing new parameters in the
form of vectors $\vec{a}$ and $\vec{b}$, not considered in
previous works and leading to linear terms in the phonon
dispersion law. The latter vector parameters are also determined
by a fitting of the bulk semiconductor dispersion law for the
optical phonons. Equation~(\ref{1})  represents a system of three
coupled partial differential equations assumed to describe the
confined polar optical phonons in each segment of the
semiconductor heterostructure. The solutions must be matched at
the interfaces by applying the boundary conditions for $\vec{u}$,
which we shall analyze below.

It is important to notice that all the four terms at the r.h.s. of
Eq. (\ref{1}) may be expressed as the divergence of a tensor
$\sigma_{ij}$, casted as

\be \label{2} \sigma_{ij}=\rho
(\beta_L^2-2\beta_T^2)(\nabla\cdot\vec{u})\delta_{ij}+\rho\beta_T^2\left(u_{j,i}+u_{i,j}\right)+i\rho\left
(b_iu_j+a_ju_i\right )\,\,.
\ee

This tensor could be compared to the stress tensor used in the
theory of continuous elastic media, and actually its divergence
provides a force density. However, in the framework of the present
approach, it should be considered as a phenomenological quantity,
which is introduced {\it ad hoc} in order to provide a description
of {\it dispersive} optical phonons in close correspondence with
what is seen both experimentally and in atomistic model
calculations. In this spirit we do not attempt to explain the tensor
$\sigma_{ij}$ in terms of the elastic properties of the system,
and tensor $\sigma_{ij}$ should not be considered a stress tensor
in the strict sense of this concept. The suffixes
$i,\,j=1,\,2,\,3$, denoting the cartesian components,  should not
be confused with the $i=\sqrt{-1}$ used as a factor. Obviously, in
all the equations above we assume an isotropic model for the
semiconductor. The short notation $u_{i,\,j}$ denotes the partial
derivative with respect to $x_j$, i. e., $u_{i,\,j} = \partial u_{i}/\partial x_{j}$.

Equations~(\ref{1}) and (\ref{2}) shall be now applied to the
nanowire case, assuming it is an infinite cylinder of circular
cross-section with radius $r_0$. The use of cylindrical
coordinates ($r,\,\theta,\,z$) becomes a natural choice, and the
equations must be written in terms of these coordinates, as well
as the tensor $\sigma_{ij}$. In the theory of elastic continuous media
the standard boundary conditions applied to any interface between two different materials are:
(1) continuity of the
displacement vector $\vec{u}$; (2) continuity of the force flux,
given by ${\bf \sigma}\cdot \vec{N}$, where $\vec{N}$ is the
normal unit vector to this surface. Solutions showing divergences
at any point of the region must also be disregarded and, in
particular, we must avoid singular solutions in the asymptotic
limits $r\to 0,\,\infty$.

It is convenient to introduce the auxiliary quantities
$\vec{\Gamma}=\nabla\times \vec{u}$ and
$\Lambda=\nabla\cdot\vec{u}$. Thus, after simple mathematical
manipulations, Eq.~(\ref{1}) is transformed into the following
equations:

\begin{eqnarray}
\left [\nabla^2+q^2_L\right ]\Lambda &=&0\,\,, \label{3}\\
\left [\nabla^2+q^2_T\right ]\vec{\Gamma}
&=&\frac{i}{\beta^2_T}\vec{a}\times\nabla\Lambda\,\,, \label{4}
\end{eqnarray}
where
\begin{eqnarray}
q_L^2&=&\frac{1}{\beta^2_L}\left
[\omega^2_0-\omega^2+i(\vec{a}+\vec{b})\cdot
\nabla\right]\,\,,\label{5}\\
q_T^2&=&\frac{1}{\beta^2_T}\left
[\omega^2_0-\omega^2+i\vec{b}\cdot \nabla\right]\,\,.\label{6}
\end{eqnarray}

Notice that the latter quantities are actually differential
operators in close correspondence with the starting equation.
Another important issue involving Eq.~(\ref{4}) concerns its
coupling to Eq.~(\ref{3}). All these features make the differences
with respect to previous treatments,\cite{r17,r18,r21} and lead
to mathematical complications as a result of our aim of
introducing linear terms in the phonon dispersion law. When
applying the equations to the case of an infinite cylinder of
radius $r_0$ filled with an isotropic non polar semiconductor, we
may make use of the translational and rotational symmetries of the
system, assuming that the solutions must be proportional to $\exp
[i(n\theta +kz)]$, where $n=0,\,1,\,2,\cdots$ and $k$ plays the
role of the wave vector. Moreover, vectors $\vec{a}$ and $\vec{b}$
shall be taken along the $z$-axis, provided that, with this
choice, we are led to the appropriate linear dependence on $k$ for
the phonon frequencies. Under such conditions $q^2_{L(T)}$ become
parameters linearly depending on the wave vector $k$ and given by:
$q^2_{L}=(\omega_0^2-\omega^2-(a+b)k)/\beta_L^2$ and
$q^2_{T}=(\omega_0^2-\omega^2-bk)/\beta_T^2$.

Another point of central importance concerns the boundary
conditions that should be applied. In this work we shall consider
the nanowire as ``free standing'', so the appropriate boundary
conditions involve zero force at the cylindrical surface of radius
$r_0$. After rewriting tensor $\sigma_{ij}$ in cylindrical
coordinates, and realizing that the normal unit vector $\vec{N}$
must be taken along the radial direction ($\vec{N}=\vec{e}_r$), we
should require the components $\sigma_{rr}$, $\sigma_{\theta r}$
and $\sigma_{zr}$ to be equal to zero at $r=r_0$. Starting from
Eq.~(\ref{2}), tediously long but straightforward mathematical
manipulations lead us to the following results:

\begin{eqnarray}
\sigma_{rr}&=&\beta_L^2\frac{\partial u_r}{\partial
r}+(\beta_L^2-2\beta_T^2)\left (\frac{1}{r}\frac{\partial
u_{\theta}}{\partial \theta}+\frac{\partial u_z}{\partial z}
+\frac{1}{r}u_r \right ) \,\,,\label{7}\\
\sigma_{\theta
r}&=&\beta_T^2\left (\frac{\partial u_{\theta}}{\partial
r}+\frac{1}{r}\frac{\partial u_r}{\partial
\theta}-\frac{1}{r}u_{\theta}\right )
\,\,,\label{8}\\
\sigma_{zr}&=&\beta_T^2\left (\frac{\partial
u_r}{\partial z}+\frac{\partial u_z}{\partial r}\right
)+ibu_r\,.\label{9}
\end{eqnarray}

\section{Solution of the differential equations}

Assuming $\Lambda (r,\,\theta,\,z)=f(r)\exp[i(n\theta+kz)]$, Eq.~(\ref{3}) leads to

\be\label{10}
\frac{d^2f}{d\xi^2}+\frac{1}{\xi}\frac{df}{d\xi}+\left
(1-\frac{n^2}{\xi^2}\right ) f=0\,, \ee where $\xi=Q_Lr$ and
$Q^2_L=q^2_L-k^2$. Eq.~(\ref{10}) is identified as Bessel equation
of order $n$, and its linearly independent (LI) solutions are usually denoted by
$J_n(\xi)$ and $Y_n(\xi)$, Bessel functions of the first and
second kind, respectively.\cite{r26} In our case we shall apply
just the function $J_n(\xi)$, convergent in the interior of the
cylinder with $0\leq r\leq r_0$.

Solutions of Eq.~(\ref{4}), an inhomogeneous Helmholtz equation
for the vector quantity $\Gamma$, are more difficult to find. In
the first place we again shall consider that
$\vec{\Gamma}(r,\,\theta,\,z)\sim \exp[i(n\theta+kz)]$. We have
found a particular solution of Eq.~(\ref{4}), which is given by

\be\label{11} \vec{\Gamma}_P=A\vec{e}_z\times\nabla\Lambda\,,\quad
\mbox{with}\quad A=\frac{ia}{\beta_T^2(q_T^2-q_L^2)}\,, \ee where
$\vec{e}_z$ is a unit vector along the $z$-axis.

Then, we just have to find the solution of the homogeneous
Helmholtz equation $\vec{\Gamma}_H$, and the general solution
shall be given by $\vec{\Gamma}=\vec{\Gamma}_P+\vec{\Gamma}_H$.
As discussed in Ref.~\onlinecite{r26}, page 1766, the two LI
solutions of the homogeneous Helmholtz equation for a vector
function $\vec{\Gamma}$, using cylindrical coordinates, are casted
as

\be\label{12}
\vec{\Gamma}_{H1}=\nabla v_1\times \vec{e}_z\, \quad
\mbox{and}\quad \vec{\Gamma}_{H2}=q_Tv_2
\vec{e}_z+\frac{1}{q_T}\nabla \frac{\partial v_2}{\partial z}\,,
\ee where the functions $v_i$  ($i=1,\,2$) satisfy the scalar
homogeneous Helmholtz equations $(\nabla^2+q_T^2)v_i=0$, with
solutions analogous to those of $\Lambda$. After the general
solutions for $\Lambda$ and $\vec{\Gamma}$ are found, we can find
the corresponding solution of Eq.~(\ref{1}) for $\vec{u}$, which
shall be obtained from

\be\label{13} \vec{u}=\frac{1}{q_T^2}\left [
\nabla\times\vec{\Gamma}-\left ( \frac{\beta_L}{\beta_T}\right
)^2\nabla \Lambda-\frac{ia}{\beta_T^2}\Lambda \vec{e}_z \right
]\,. \ee

In the foregoing calculations the involved mathematical steps are
long and tedious, but here we shall report just final expressions.
We write $\vec{u}=\vec{F}(r)\exp[i(n\theta +kz)]$ where
$\vec{F}=F_r\vec{e}_r+F_{\theta}\vec{e}_{\theta}+F_z\vec{e}_z$ and
$\vec{e}_r$, $\vec{e}_{\theta}$, and $\vec{e}_z$, are the unit
vectors of the cylindrical coordinates. Then, we have

\begin{eqnarray}
F_r(r)&=&J\p_n(Q_Lr)C_1+\frac{ik}{Q_T}J\p_n(Q_Tr)C_2+\frac{in}{Q_Tr}J_n(Q_Tr)C_3\,,\label{14}\\
F_{\theta}(r)&=&\frac{in}{Q_Lr}J_n(Q_Lr)C_1-\frac{nk}{Q^2_Tr}J_n(Q_Tr)C_2-J\p_n(Q_Tr)C_3\,,\label{15}\\
F_z(r)&=&\frac{i(k-\gamma)}{Q_L}J_n(Q_Lr)C_1+J_n(Q_Tr)C_2\label{16}\,,
\end{eqnarray}
where $Q_T^2=q_T^2-k^2$ and  $\gamma =a/(\beta_T^2-\beta_L^2)$.
The ``prime'' in the Bessel function denotes the first derivative
with respect to the function's argument and the constants $C_i$
($i=1,\,2,\,3$) should be determined after application of the
boundary conditions. As it was already remarked in Section II, we
consider a ``free standing'' wire, and the boundary conditions
read as

\be\label{17} \sigma_{rr}|_{r_0}=0\,,\quad \sigma_{\theta
r}|_{r_0}=0,\,\quad \sigma_{zr}|_{r_0}=0\,\,, \ee
indicating that
the internal flux of forces through the boundary surface is zero.
Taking Eqs.~(\ref{7}), (\ref{8}),  (\ref{9}), and
(\ref{17}) into account we are lead to the following system of linear
homogeneous equations

\be\label{18} \sum_{j=1}^3G_{ij}C_j=0\,, \ee
where the nine
coefficients $G_{ij}$ are explicitly given in the Appendix. The
dispersion relations, determining the frequencies of the various
phonon modes as  functions of the wave vector $k$, are obtained
after solution of the secular equation

\be \label{19}
\det\left [G_{ij}\right ] =0\,,
\ee
while the components of the vector $\vec{F}$ are given as

\begin{eqnarray}
F_r(r)/C&=&S_1J\p_n(Q_Lr)+\frac{ik}{Q_T}S_2J\p_n(Q_Tr)+\frac{in}{Q_Tr}S_3J_n(Q_Tr)\,,\label{20}\\
F_{\theta}(r)/C&=&\frac{in}{Q_Lr}S_1J_n(Q_Lr)-\frac{nk}{Q_T}S_2\frac{1}{Q_Tr}J_n(Q_Tr)-S_3J\p_n(Q_Tr)\,,\label{21}\\
F_z(r)/C&=&\frac{i(k-\gamma)}{Q_L}S_1J_n(Q_Lr)+S_2J_n(Q_Tr)\,,\label{22}
\end{eqnarray}
where the functions $S_i$ ($i=1,\,2,\,3$) are given in the
Appendix.

\section{Normalization of the phonon states}

The determination of the constant $C$ is made by normalization of
the oscillation modes, which requires to transform the classical
field $\vec{u}$ into a quantum-field operator. This is formally done
in the form $\vec{u}\to \hat{\vec{u}}=\vec{u}\hat{a}_{n,\,k}$, where
$\hat{a}_{n,\,k}$ are second quantization bosonic operators.
Operator $\hat{a}_{n,\,k}$ ($\hat{a}_{n,\,k}^{\dagger}$) annihilates
(creates) a ``$n,\,k$'' optical phonon. We must also transform the
classical kinetic energy for the oscillations to a
quantum-mechanical operator by applying the rule

\be \label{29}
W_{kin}=\frac{1}{2}\rho\int_V
\dot{\vec{u}}^{*}\cdot\dot{\vec{u}}\,d^3r=\frac{1}{2}\rho\omega^2\int_V
\vec{u}^{*}\cdot{\vec{u}}\,d^3r\;\to \hat{H}_{ph}\;, \ee where \be
\label{30} \hat{H}_{ph}=\frac{1}{4}\rho\omega^2\int_V \left
(\hat{\vec{u}}^{\dagger}\cdot\hat{\vec{u}}+\hat{\vec{u}}\cdot\hat{\vec{u}}^{\dagger}\right
)\,d^3r=\frac{1}{2}\rho\omega^2\int_V\vec{u}^{*}\cdot\vec{u}\,d^3r\,\left
(\hat{a}_{n,\,k}^{\dagger}\hat{a}_{n\,,k}+\frac{1}{2}\right )\;.
\ee

In Eq.~(\ref{30}) $\hat{H}_{ph}$ describes the free phonons
hamiltonian operator, while its hermitian character is ensured by
construction. Requiring the latter expression to be given in
standard form for phonons of the type ``$n\,,k$'' (actually, we
must have $\omega \to \omega_{n\,k}$)
$\hat{H}_{ph}=\hbar\omega_{n\,,k}\left
(\hat{a}_{n\,,k}^{\dagger}\hat{a}_{n,\,k}+\frac{1}{2}\right )$,
one can finally write the normalization constant $C$ by

\be \label{31} C_{n,\,k}=\sqrt{\frac{\hbar}{\pi
L\rho\omega_{n,\,k}M_{n,\,k}}}\;, \ee
where $L$ is a normalization
length (taken along the $z$ axis) and $M_{n,\,k}=\int
(F_r^{*}(r)F_r(r)+F_{\theta}^{*}(r)F_{\theta}(r)+F_z^{*}(r)F_z(r))rdr/|C_{n,\,k}|^2$.
The explicit expressions for $M_{n,\,k}$ involve rather
complicated integrations of the Bessel functions and may be
calculated numerically for each value of $k$ and $\omega$
corresponding to the possible oscillation modes.

For optical phonons in a non polar semiconductor the
electron-phonon hamiltonian is given through the deformation
potential approximation and, for an isotropic semiconductor, is
proportional to $\nabla \cdot \vec{u}$. Then, it is interesting to
discuss this latter quantity, which provides an estimative of the
interaction strength.

\section{Discussion of results}

With the aim of illustrating the foregoing theory, we consider a Ge
nanowire. The Ge physical parameters are given in Table I.

\begin{table}[h]
\caption{Ge parameters}
\begin{center}
\begin{tabular}{|c|c|c|c|c|c|c|c|}
\hline
material & $a_0[10^{-8}cm]$ & $\rho\,[g/cm^3]$ &
$\omega_0\,[cm^{-1}]$  & $\beta_T\,[10^5cm/s]$ &
$\beta_L\,[10^5cm/s]$&$a[10^{18}cm/s^2]$&$b[10^{18}cm/s^2]$ \\
\hline $Ge$ &5.66 &5.32 & 305 & 3.48  &3.55  &-6.26&3.59\\ \hline
\end{tabular}\\
\end{center}
\end{table}
The parameters $\beta_T$, $\beta_L$, $a$ and $b$ were determined by
the authors by a fitting of the bulk semiconductor phonon
dispersion law reported in Ref.\onlinecite{r27}, considering
dispersion along the $(001)$ direction and embracing approximately
$50-40\;\;\%$ of the BZ near the $\Gamma$ point. Parameter
$\omega_0$ was taken from the same reference, while the other
parameters can be found in Ref.~\onlinecite{r28}.

\begin{figure*}[tbp]
\includegraphics*{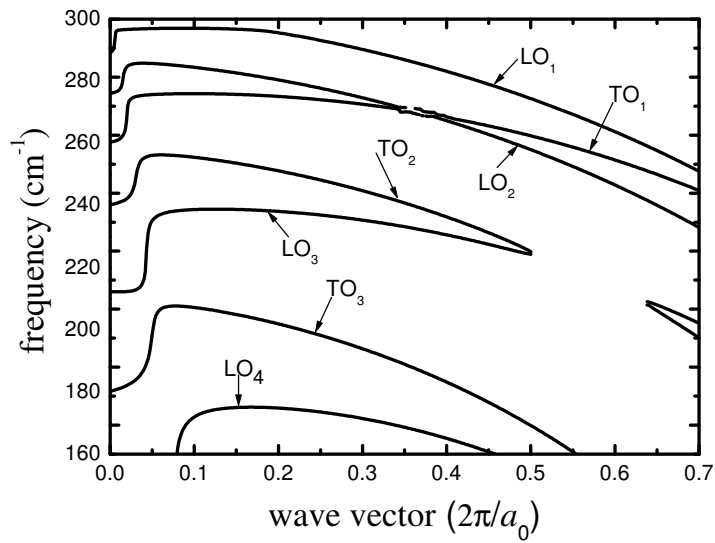}
\caption{Dispersion curves for the axis-symmetric
modes (n=0). The frequency is in units of cm$^{-1}$ and the wave vector $k$ in units of
$2\pi /a_0$) and we take $r_0=10^{-7}$ cm. \label{1}}
\end{figure*}

In Fig.~1 we depict the dispersion relation for several axis-symmetric modes
($n=0$) taking $r_0=10^{-7}$ cm. The frequency is in units of cm$^{-1}$)
and the wave vector $k$ is in units of $2\pi/a_0$, where
$a_0$ is the lattice parameter for the bulk semiconductor). The
curves were obtained from Eq.~(\ref{19}) taking $n=0$. Let
us emphasize that these modes do not present torsional
oscillations, and they result from a combination of radial
breathing vibrations together with vibrations along the cylinder
axis (axial oscillations). The latter oscillations are not purely
longitudinal as it is sometimes argued, but a mixture of LO and TO
oscillations. However, at $k =0$ the axis-symmetric oscillations
separate into pure LO and TO vibrations, a fact we have used for
the labelling of the different modes.

\begin{figure*}[tbp]
\includegraphics*{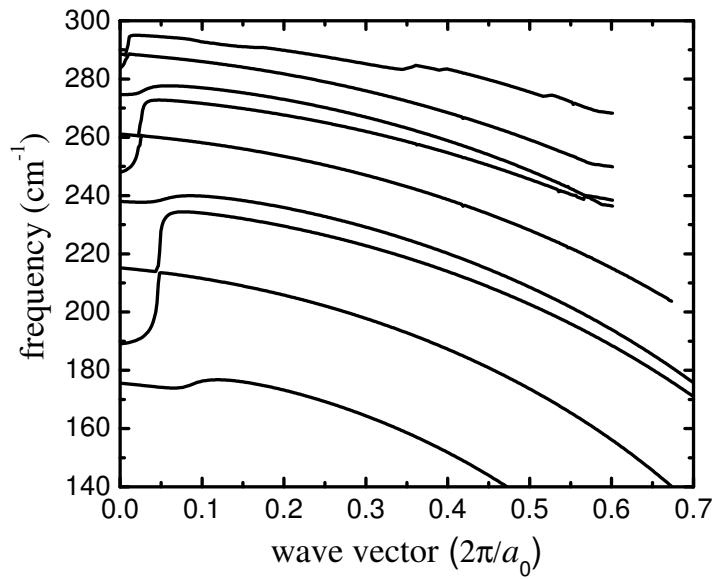}
\caption{Dispersion curves for the flexural modes for $n=1$ taking $r_0=10^{-7}$ cm.
The frequency is in units of cm$^{-1}$) and the wave vector $k$ in units of $2\pi /a_0$). \label{2}}
\end{figure*}

Notice that for the curves corresponding to the modes TO$_2$ and
LO$_3$  there is an interval of $k$ values (approximately between
0.5 and 0.63) where the curves are interrupted, and no frequencies
are reported. This region obviously corresponds to an anticrossing
of the mentioned modes. Of course, in the real case, there must be
oscillation frequencies and the interrupted curves may be
interpreted as a limitation of our approach, which is not able to
describe such oscillation modes. However, it is worth to mention
that the observed failure has nothing to due with the anticrossing
effect by itself. The theory we are applying is essentially valid
not too far from the $k =0$ region, and we are actually
extrapolating it beyond this region. This latter issue should be
seen as a singular failure of the applied treatment, which,
otherwise is giving a rather good description of the dispersion
law for these modes. We have set $r_0=10^{-7}$ cm, a relatively
small radius which, however, is typical for this kind of
nanowires. For such thin nanowire the number of physically
possible optical phonon modes is rather limited and, in fact,
Fig.~1 displays a few more modes than actually present. In the
framework of the current paper one important contribution is to
stress the importance of taking the role of the
linear term into account we have added to our equations. It is easy to convince
ourselves of this latter issue: making the linear terms equal to
zero the general structure of the dispersion law depicted in
Fig.~1 is subjected to significant changes. The other interesting
point is to compare with experimental results or calculations
based upon microscopic (atomistic) approaches. Unfortunately, in
the revised literature we have not found reports on optical
phonons for this particular type of Ge nanowires.

\begin{figure*}[tbp]
\includegraphics*{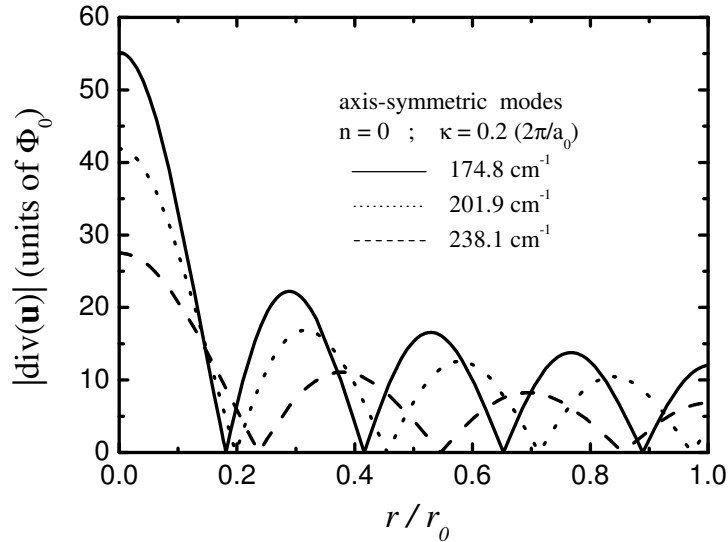}
\caption{Absolute value of the divergence of vector $\vec{u}$ (in
units of $\Phi_0$) as a function of $r/r_0$ for the axis-symmetric
modes ($n=0$). We fixed $k = 0.2 (2\pi /a_0)$ and considered the
frequencies $174.8$, $201.9$ and $238.1$ cm$^{-1}$ corresponding
to three possible modes displayed in Fig.~1. \label{3}}
\end{figure*}

In Fig.~2 we show the dispersion curves for the so-called
flexural modes taking again $r_0=10^{-7}$ cm and for $n=1$. These
modes involve all kinds of possible vibrations: radial, axial and
torsional. Torsional modes can be considered for any $n>0$ and we
are analyzing the case with $n=1$. In the chosen frequency interval
($140 <\omega < 300$ cm$^{-1}$), involving the higher frequency
values, just nine modes are present. We should remark that, for the
flexural modes, the oscillations have always a mixed character
and cannot be separated into pure LO and TO oscillations. In
contrast with the axis-symmetric modes the mixing is present even at
$k = 0$, and then we cannot classify the modes in the same way as we did in
Fig.~1. Other feature clearly seen in Fig.~2 is the presence of
several anticrossing regions: curves belonging to the same symmetry
cannot cross each other and are ``repelled''  whenever they get too
close. Finally, there are several curves (those with the higher
frequencies) that are interrupted for the higher values of $k$
(especially, in the region $k > 0.6$, in units of $2\pi /a_0$). The
explanation for this effect is the same as in Fig.~1: a failure of
the used approximation, which is actually valid for values of $k $
not too far from $k =0$. However, we must again remark that the
theory actually gives a very useful description of the
considered modes for a relatively wide interval of $k$ ($60 -
70\,\,\%$ of the BZ).

\begin{figure*}[tbp]
\includegraphics*{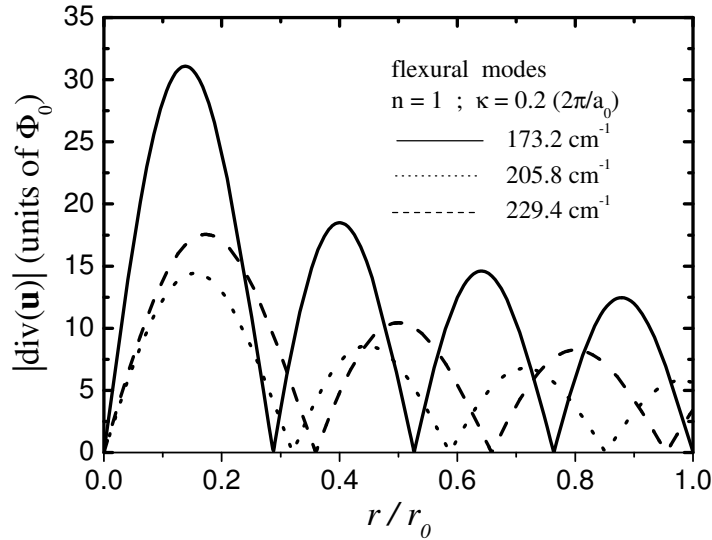}
\caption{Absolute value of the divergence of vector $\vec{u}$ (in
units of $\Phi_0$) as a function of $r/r_0$ for the flexural modes
($n=1$). We fixed $k = 0.2 (2\pi /a_0)$ and considered the
frequencies $173.18$, $205.8$ and $229.4$ cm$^{-1}$ corresponding
to three possible modes displayed in Fig.~2. \label{4}}
\end{figure*}

Besides the dispersion curves, we find interesting to
analyze the quantity given by $\nabla\cdot\vec{u}\equiv
div\,(\vec{u})$, as a function of $r$ in the interval $r\leq r_0$.
The point is that this quantity is proportional to the
electron-phonon interaction according to the {\it deformation
potential} approximation, and then can give us an estimative of
the strength of this interaction. We have taken the divergence of
vector $\vec{u}$, given in Eq.~(\ref{22}) in cylindrical
coordinates, together with the normalization constant,
Eq.~(\ref{31}), which was evaluated numerically.

In Fig.~3 we plot
the absolute value of the mentioned quantity ($|div\,( \vec{u})|$)
in units of $\Phi_0=[\hbar /L\rho\pi\omega_0r_0^4]^{(1/2)}$ as a
function of $r/r_0$. We show three curves corresponding to the
axis-symmetric modes ($n=0$) with a fixed value of $k$ ($k =0.2
(2\pi /a_0$) and three possible frequencies ($\omega =
174.8,\;201.9,\;238.1$ cm$^{-1}$) involving three different modes
(the modes LO$_4$, TO$_3$, and L$O_3$ respectively). In Fig.~3 the
oscillatory behavior of the interaction potential strength is
seen, a clearly understandable fact if we realize that it is given
as a combination of Bessel functions. At the point $r=0$  the
quantity is not zero, while the points of maxima determine the
values of $r$ where the interaction strength should be stronger.
Moreover, for the modes with increasing values of the frequency the
strength of the interaction gets weaker. When plotting the
absolute value of $div\,(\vec{u})$ we have lost all information
concerning its dependence with respect to $\theta$ and $z$.
However, we retained the important information about its $r$
dependence. In should also be kept in mind that, in order to
obtain a better estimation of the electron-phonon interaction
strength, it is necessary to evaluate the corresponding matrix
elements involving the electron wave functions. If the electron
wave function is peaked at a point where the deformation potential
has a low value, we should realize that the interaction strength
still may be strong depending on the relative weight of both
effects. The plot displayed in Fig.~4 is quite similar to that of
Fig.~3, but now considering the $n=1$ case (flexural modes). We
again see the oscillatory behavior for the same reason as in
Fig.~3, but now the involved quantity is zero at $r=0$, and the
calculated strengths are somewhat lower. It is also seen that the
deformation potential is not lower for higher values of the
frequency. In fact, the intermediate frequency mode leads to
weaker intensities that the lower frequency mode. The same
comments as those done for Fig.~3 can be made.

In conclusion, we have calculated the optical phonon modes for a
Ge nanowire in the framework of a continuum approach. The applied
theoretical treatment was adapted to the non polar case, and also
was improved by incorporating the important linear terms in the
wave vector. In this context we expect the theory to provide better
results than those given by other long wavelength treatments. But,
as was already remarked, we do not presently have experimental
results or microscopic atomistic calculations allowing a desirable
comparison with our findings. Nevertheless, and due to the current
interest in the study of this kind of nanostructures, we think the
results of the present work should be useful in the way to get a
better understanding of Ge and Si nanowires.

\appendix

The nine functions $G_{ij}$, appearing in Eq. (\ref{18}), are given in the form

\begin{eqnarray}\
G_{11}&=&\beta_L^2Q_L\left\{ \left [ \left (\frac{\beta_Tn}{\beta_L\eta}\right )^2-\left (1-\left (\frac{\beta_T}{\beta_L}\right )^2\right )\nu (\nu-\alpha)/\eta^2-1\right ]J_n(\eta)-2\left (\frac{\beta_T}{\beta_L}\right )^2J\p_n(\eta)/\eta\right \}\,,\nonumber\\
G_{12}&=&\frac{2ik\beta_T^2}{\mu^2}\left [(n^2-\mu^2)J_n(\mu )-\mu J_n\p(\mu )\right ]\,,\nonumber\\
G_{13}&=&-\frac{2in\beta_T^2Q_T}{\mu^2}\left [J_n(\mu )-\mu J_n\p(\mu )\right ]\,,\nonumber\\
G_{21}&=&-\frac{2in\beta_T^2Q_L}{\eta^2}\left [J_n(\eta )-\eta J_n\p(\eta )\right ]\,,\nonumber\\
G_{22}&=&\frac{2nk\beta_T^2}{\mu^2}\left [J_n(\mu )-\mu J_n\p(\mu )\right ]\,,\nonumber\\
G_{23}&=&\frac{\beta_T^2Q_T}{\mu^2}\left [2\mu J_n\p(\mu )+(\mu^2-2n^2)J_n(\mu )\right]\,,\nonumber\\
G_{31}&=&i\beta^2_Tk\left [ 2-\alpha/\nu +br_0/(\beta^2_T\nu )\right ]J_n\p(\eta )\,,\nonumber \\
G_{32}&=&-\frac{\beta_T^2k^2}{Q_T}\left [1-(\mu/\nu )^2+br_0/(\beta_T^2\nu)\right ]J_N\p(\mu )\,,\nonumber\\
G_{33}&=&-\frac{nk\beta_T^2}{\mu}\left [1+br_0/(\beta_T^2\nu)\right ]J_n(\mu )\,,
\end{eqnarray}
where $\alpha=\gamma r_0$, $\nu=kr_0$, $\eta=Q_Lr_0$, and $\mu=Q_Tr_0$. On the other hand,
the functions $S_i$, present in Eqs.~(\ref{20}), (\ref{21}) and (\ref{22}),
are defined as

\begin{eqnarray}
S_1&=&G_{32}G_{23}-G_{22}G_{33}\,,\nonumber\\
S_2&=&G_{21}G_{33}-G_{31}G_{23}\,,\nonumber\\
S_3&=&G_{31}G_{22}-G_{32}G_{21}\,.\nonumber\\
\end{eqnarray}

\begin{acknowledgments} The work is partially supported by
Funda\c{c}\~ao de Amparo \`a Pesquisa de S\~ao Paulo and
Conselho Nacional de Desenvolvimento Cient\'{i}fico e
Tecnol\'{o}gico. F.C. is grateful to Departamento de
F\'{\i}sica, Universidade Federal de S\~ao Carlos, for
hospitality.
\end{acknowledgments}


\begin{references}
\bibitem{rr1} In this formula $\omega$ is the frequency (in units of $rad/s$),
$k$ is the wave vector, $\beta$ a phenomenological parameter used to fit the experimentally reported bulk semiconductor dispersion law for the phonons near the point $k=0$ of the BZ. Obviously, $\omega_0$ is the phonon frequency at $k=0$.
\end{references}
\end{document}